\documentclass[aps,prb,twocolumn,amsmath,amssymb,superscriptaddress]{revtex4-1}
\usepackage[english]{babel}
\usepackage[utf8]{inputenc}
\usepackage{graphicx}
\usepackage{xfrac}
\usepackage{amsfonts}
\newcommand{\LAO}{LaAlO$_3$}
\newcommand{\STO}{SrTiO$_3$}

\newcommand{\dgC}{$\,^{\circ}\mathrm{C}$}
\newcommand{\SrO}{SrO}
\newcommand{\TiO}{TiO$_2$}
\newcommand{\LaO}{LaO}
\newcommand{\AlO}{AlO$_2$}

\newcommand{\ie}{i.e.\@}
\newcommand{\surfAl}{Al$_{3/2}$O$_2$}
\newcommand{\surfLa}{La$_{5/6}$O}

\newcommand{\SampleCondA}{S1-C}
\newcommand{\SampleOutOfFocus}{S2-I}
\newcommand{\SampleTwente}{S4-C}
\newcommand{\SampleSrO}{S3-I}
\newcommand{\SampleSputtered}{S5-I}

\newcommand{\SampleCond}{\SampleCondA}
\newcommand{\SampleIns}{\SampleOutOfFocus}

\newcommand{\SampleOscillationsCond}{\SampleCondA}
\newcommand{\SampleOscillationsIns}{\SampleOutOfFocus}

\newcommand{\ElectricARRESCond}{blue}
\newcommand{\ElectricTwente}{red}
\newcommand{\ElectricOutOfFocus}{green}
\newcommand{\ElectricSrO}{cyan}
\newcommand{\ElectricSputtered}{magenta}

\newcommand{\ElectricCond}{\ElectricARRESCond}
\newcommand{\ElectricIns}{\ElectricOutOfFocus}

\newcommand{\SampleARRESCond}{\SampleCondA}

\newcommand{\SampleARRESIns}{\SampleOutOfFocus}

\begin{document}

\title{Formation of a conducting \LAO{}/\STO{} interface studied by low energy electron reflection during growth }
%
\author{A.J.H. van der Torren}
\affiliation{Huygens - Kamerlingh Onnes  Laboratorium, Leiden University, Niels Bohrweg 2, 2300 RA Leiden, The
Netherlands}
\author{Z. Liao}
\affiliation{MESA+ Institute for Nanotechnology, University of Twente, PO Box 217, 7500 AE Enschede, The Netherlands}
\author{C. Xu}
\altaffiliation[New address: ] {Institut für Solarenergieforschung GmbH Hameln/Emmerthal, Am Ohrberg 1, 31860
Emmerthal, Germany}
\affiliation{Peter Gr\"{u}nberg Institute, Forachungscenter J\"{u}lich, 52425, J\"{u}lich, Germany}
\author{N. Gauquelin}
\affiliation{EMAT, University Antwerp, Groenenborgerlaan 171, BE-2020 Antwerp, Belgium}
\author{C. Yin}
\affiliation{Huygens - Kamerlingh Onnes  Laboratorium, Leiden University, Niels Bohrweg 2, 2300 RA Leiden, The
Netherlands}
\author{J. Aarts}
\email[email: ]{aarts@physics.leidenuniv.nl} \affiliation{Huygens - Kamerlingh Onnes  Laboratorium, Leiden University,
Niels Bohrweg 2, 2300 RA Leiden, The Netherlands}
\author{S. J. van der Molen}
\affiliation{Huygens - Kamerlingh Onnes  Laboratorium, Leiden University, Niels Bohrweg 2, 2300 RA Leiden, The
Netherlands}

\date{\today}

\begin{abstract}
The two-dimensional electron gas occurring between the band insulators \STO{} and \LAO{} continues to attract
considerable interest, due to the possibility of dynamic control over the carrier density, and the ensuing phenomena
such as magnetism and superconductivity. The formation of this conducting interface is sensitive to the growth
conditions, but despite numerous investigations, there are still questions about the details of the physics involved.
In particular, not much is known about the electronic structure of the growing \LAO{} layer at the growth temperature
(around 800\dgC{}) in oxygen (pressure around $5\times10^{-5}$~mbar), since analysis techniques at these conditions are
not readily available. We developed a pulsed laser deposition system inside a low-energy electron microscope in order
to study this issue. The setup allows for layer-by-layer growth control and in-situ measurements of the angle-dependent
electron reflection intensity, which can be used as a fingerprint of the electronic structure of the surface layers
{\it during} growth. By using different substrate terminations and growth conditions we observe two families of
reflectivity maps, which we can connect either to samples with an AlO$_2$-rich surface and a conducting interface; or
to samples with a LaO-rich surface and an insulating interface. Our observations emphasize that substrate termination
and stoichiometry determine the electronic structure of the growing layer, and thereby the conductance of the
interface.
\end{abstract}


\maketitle
\newpage

\section{Introduction}
Transition metal oxides, and in particular perovskites, form an important class of materials exhibiting a variety of
physical phenomena such as superconductivity, magnetism and ferroelectricity. Especially interesting for possible
electronics applications is the occurrence of a two-dimensional electron gas between the two band insulators \LAO{} and
\STO{}~\cite{ohtomo04}. The emergence of this conducting interface can at least partially be explained by the so-called
polar catastrophe model. In this model an increasing electrical potential builds up when charged (LaO)$^+$ and
(\AlO{})$^-$ layers are alternatively stacked on top of neutral layers of SrO and TiO$_2$. This potential is
compensated by the transfer of half an electron charge from the surface to the interface. A relevant observation is
that the electron gas only forms when the top \LAO{} layer is at least four unit cells thick~\cite{thiel06}. At that
thickness the potential buildup is apparently enough to transfer the charge to the interface. Furthermore, the electron
gas only forms at the n-type interface (\TiO{}/\AlO{}) and not at the p-type interface (\SrO{}/\LaO{})~\cite{ohtomo04}.
At the p-type interface a structural reconstruction is energetically favored over the electronic
reconstruction~\cite{zhang10}.

Other observations, however, are at odds with a simple electronic reconstruction model. To name just two, electrical
field build-up in the \LAO{} layer below the critical thickness is not observed~\cite{segal09,slooten13}, and samples
grown in high oxygen partial pressure do not conduct~\cite{herranz07,kalabu11}. Clearly, defects in the \LAO{} layer
and in the \TiO{} termination layer also play an important role in the formation of the electron gas. Not surprisingly,
therefore, it is very much the growth conditions which determine the conducting properties of the interface. In Pulsed
laser deposition (PLD), the exact plume shape and composition as well as the oxygen pressure are of great importance,
influencing the cationic stoichiometry\cite{warusa13,brecken13a} of the \LAO{} film and the number of oxygen vacancies
in the \STO{}~\cite{brinkman07}. In particular, a La/Al-ratio exceeding 0.97~\cite{warusa13} was shown to fully
suppress the conductivity. Similarly, high-pressure oxygen sputtering yielded a La/Al ratio well above 1 and
non-conducting interfaces~\cite{dildar13}. Also the occurrence of magnetism~\cite{brinkman07, ariando11, dikin11,
bert11} and superconductivity~\cite{reyren07, joshua12, caviglia08} was shown to be sensitive to the oxygen pressure
during growth.
\begin{figure*}[t]
\includegraphics[width=14cm]{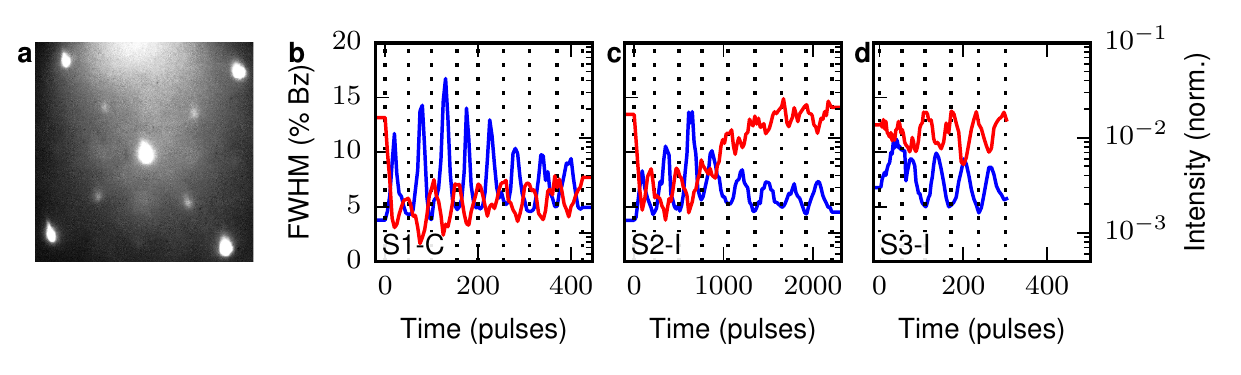} \caption{ a) Diffraction pattern on bare \STO{} at a growth temperature of
820~\dgC{} taken at 17~eV. b) FWHM (blue) and maximum intensity (red) of the specular diffraction spot for a conducting
sample~\SampleOscillationsCond{}. c) Same for the insulating sample~\SampleOscillationsIns{} and d) same for the
insulating sample~\SampleSrO{}. All data have been taken at 17~eV landing energy. The FWHM is given in percentage of
the Brillouin zone, which is equal to the percentage of the distance from specular to the first order spots. The
intensity has been normalized at the mirror mode (zero landing energy) intensity.
    }
\label{fig:FWHM}
\end{figure*}

Whereas differences in growth conditions are therefore known to lead to conducting or insulating samples as measured
afterwards, little is known about how the electronic properties of the material develop during growth, mainly because
the high temperatures and high oxygen pressure required during growth limit the abilities for in-situ analysis. For
this reason we recently developed an in-situ pulsed laser deposition system inside a low-energy electron microscope.
This allows us to follow the growth by monitoring oscillations in the width and intensity of the specular
beam~\cite{torren16}. At the same time, it allows measurements of the angle-dependent electron reflectivity of the
surface with sub-unit cell precision, which yields information on the unoccupied part of the band
structure~\cite{jobst15,jobst16}. Here we show the results of the growth of \LAO{} on \STO{} under different
circumstances. We find clear differences in the development of the reflectivity maps when growing samples with
conducting or with insulating interfaces, and relate that to the surface termination and stiochiometry of the growing
film.

\section{Experimental setup and sample preparation}
The \LAO{}/\STO{} interfaces are grown and studied in an aberration corrected low-energy electron microscope (LEEM) at
Leiden university, called ESCHER~\cite{tromp10,schramm11,schramm12,tromp13}. The LEEM technique has been used before to
study \STO{}~\cite{hessel14} and \LAO{}~\cite{torren15} separately. We now also developed a pulsed laser deposition
(PLD) system inside the system to allow for analysis during growth, which was already used to study the growth of
\STO{} on \STO{}~\cite{torren16}. In order to study growth, pulsed deposition is performed alternatingly with LEEM
imaging. In more detail, between every few laser / deposition pulses, the LEEM is turned on (meaning the high voltage
between objective lens and sample, required for the low-energy electrons, is switched on) and diffraction images are
obtained. From the diffraction images the intensity and shape of the specular diffraction spot is determined to monitor
the growth. After this measurement the high voltage is turned off and deposition can continue. For growth monitoring we
obtain the full-width-half-maximum (FWHM) and the peak intensity of the specular spot. In a layer-by-layer growth mode,
both the FWHM and the intensity oscillate, out of phase with one another, and allow precise control over the
deposition. To obtain a fingerprint of the unoccupied band structure, angle-resolved reflected electron spectroscopy
(ARRES) is also performed~\cite{jobst15,jobst16}. In this technique the electron reflection is measured depending on
energy and on the in-plane wave vector which is controlled by the angle of incidence of the electron beam. ARRES
utilizes the fact that the electron reflectivity strongly depends on the electron landing energy $E_0$ and the in-plane
momentum $k_\parallel$. In particular the electron reflection is low if the material has a band at the specific ($E_0$,
$k_\parallel$) of the electron so that it can couple into the band. In contrast, when ($E_0$, $k_\parallel$) of the
electron coincide with a band gap the electron reflectivity is high. Hence the "reflected-electron" or ARRES map shows
a fingerprint of the unoccupied band structure. For the ARRES measurements we obtain the total (integrated) spot
intensity which is independent of the surface roughness \ie{} the total intensity stays constant when the surface
roughens since the spot broadening lowers the maximum.
\begin{figure*}[t]
\includegraphics[width=\textwidth]{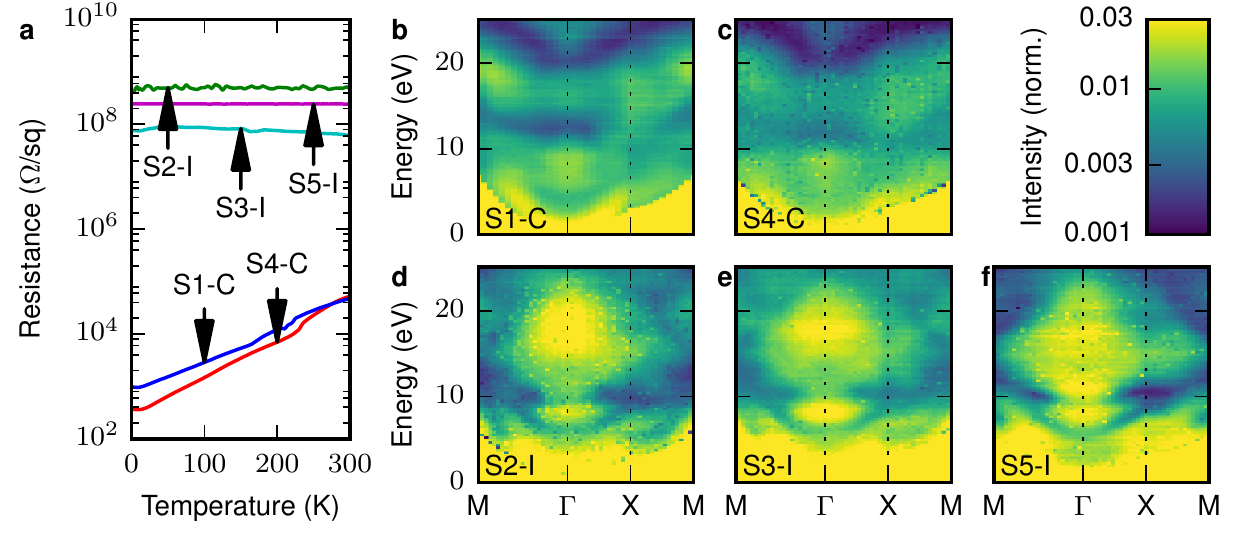}
\caption{ a) Sheet resistance versus temperature for five different samples. b-f) ARRES measurements for conducting
(\SampleCond{}, \SampleTwente{}) and non-conducting (\SampleIns{}, \SampleSrO{}, \SampleSputtered{}) samples.
Sample~\SampleCond{} (b,~\ElectricARRESCond{}) 8~u.c.\@ \LAO{} grown in the LEEM, sample~\SampleTwente{}
(c,~\ElectricTwente{}) 4~u.c.\@ \LAO{} grown in a conventional PLD setup, sample~\SampleIns{}
(d,~\ElectricOutOfFocus{}) 8~u.c.\@ \LAO{} grown with out-of-focus PLD laser, sample~\SampleSrO{} (e,~\ElectricSrO{})
5~u.c.\@ \LAO{} grown on \SrO{}-terminated \STO{} and sample~\SampleSputtered{} (f,~\ElectricSputtered) 5~nm \LAO{}
grown with sputter deposition. } \label{fig:ARREScompare}
\end{figure*}

As substrates, \STO{} (100) single crystals from CrysTec GmbH are used which were \TiO{}-terminated by a buffered HF
etch~\cite{kawasaki94} and annealing in oxygen at 950\dgC{} for one hour. The \SrO{}-terminated substrate was prepared
in a different PLD system by growing a double SrO-layer on a \TiO{}-terminated substrate. For the PLD targets, single
crystals \LAO{} (100) from Crystal GmbH were used. The PLD growth is performed at a pressure of
$5.5\times{}10^{-5}$~mbar oxygen and if not otherwise stated at a 2~J/cm$^2$ laser fluence with 1~Hz repetition rate.
Depending on deposition speed, the deposition is briefly intermitted each 5 to 50 pulses to perform imaging and
spectroscopy. This results in around 10 measurements per unit cell grown. Samples are grown at temperatures between 800
and 860 \dgC{} as measured with a pyrometer (emissivity 0.8). Temperature-dependent resistance measurements were
performed in a Physical Properties Measurement System (PPMS, Quantum Design) in a van~der~Pauw configuration. In order
to facilitate the discussion, samples with a conducting interface will henceforth be designated with the suffix "C",
insulating samples will be labeled "I".
The composition across the interface was measured by scanning transmission electron microscopy (STEM) in electron
energy loss (EELS) mode on a Titan microscope operated at 300~kV. Samples were prepared by FIB milling as described
elsewhere~\cite{chen15}. The profiles shown below (Fig.~5b) result from the average of 5 different measurements. The
integrated intensity of the Sr L-edge, Ti L edge, La M-edge and Al K -edge was normalized by dividing by the maximum. A
slight cation deficiency was ignored due to the limited precision of EELS quantification. (as discussed in
ref.~\cite{gauq14})

\section{Results}

Three \LAO{}/\STO{} heterostructures were grown under two different growth conditions and on two kind of substrates.
The first sample (\SampleCond{}) was grown with an optimal fluence of 2~J/cm$^2$ on a \TiO{}-terminated
\STO{}-substrate, the second sample (\SampleIns{}) was grown with a much lower fluence by defocusing the PLD laser on
the same \TiO{}-terminated substrate, and the third sample (\SampleSrO{}) was grown with the optimal fluence of
2~J/cm$^2$ on the \SrO{}-terminated \STO{}-substrate. For layer-by-layer growth control we took low-energy electron
diffraction images as shown in Fig.~\ref{fig:FWHM}a for bare \STO{}. The starting surface shows clear diffraction spots
and a 2~$\times$~1 reconstruction, in line with earlier observations~\cite{hessel14}. From the diffraction images, the
peak intensity and full-width-half-maximum (FWHM) of the specular spot were recorded and are shown in
Fig.~\ref{fig:FWHM}b,~c,~d in red and blue respectively for samples~\SampleCond{}, \SampleIns{} and \SampleSrO{}.

Clear oscillations can be observed in both FWHM and peak intensity, which are out of phase with one another. The
landing energy of the electrons (17~eV) has been optimized for maximal contrast in the oscillations. This energy is
close to the out-of-phase condition where the electrons destructively interfere at the step edges on the surface. At
the flat surface the coherent areas are large, resulting in sharp diffraction spots with small FWHM. When the surface
roughens during growth the coherent areas become small due to the large amounts of newly grown islands, which results
in low peak intensity and high FWHM. As a guide to the eye, dotted lines are plotted to indicate integer number of unit
cells grown. A total of eight unit cells was grown on \SampleCond{}, \SampleIns{} and five unit cells on \SampleSrO{}.
Much more pulses were needed for sample~\SampleIns{} (Fig.~\ref{fig:FWHM}c) than sample~\SampleCond{} and~\SampleSrO{}
(Fig.~\ref{fig:FWHM}b, d). From this we can conclude that the growth speed is highly reduced for the out-of-focus laser
beam, as expected.

For sample~\SampleCond{} (Fig.~\ref{fig:FWHM}b) the peak intensity strongly decreases at the start to oscillate around
a constant background for the remainder of the time. Sample~\SampleIns{} shows the same decrease of background
intensity up to two unit cells, but then comes back to the starting value between three and five unit cells.
Sample~\SampleSrO{} does not show the decrease at the start and keeps oscillating around a constant value. This change
in background intensity is related to the electronic structure of the surface layer as will become clear below. First
we characterize the electrical properties of these samples. For this, the temperature dependence of the sheet
resistance was measured and is shown in Fig.~\ref{fig:ARREScompare}a for sample~\SampleCond{} (\ElectricCond{}),
sample~\SampleIns{} (\ElectricIns{}) and sample~\SampleSrO{} (\ElectricSrO{}). Sample~\SampleCond{} shows conducting
behavior while sample~\SampleIns{} and sample~\SampleSrO{} are insulating.

\begin{figure*}[t]
\includegraphics[width=\textwidth]{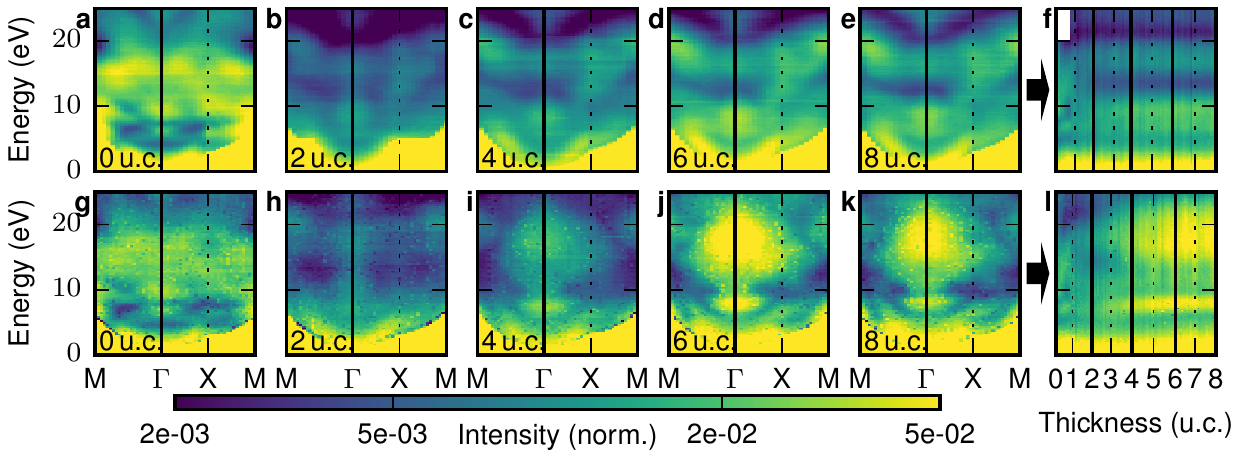}
\caption{ Conducting sample~\SampleARRESCond{} (top a-f) and non-conducting sample~\SampleARRESIns{} (bottom, g-l).
From left to right ARRES maps for 0~(a,~g), 2~(b,~h), 4~(c,~i), 6~(d,~j) and 8~(e,~k) unit cells respectively and an
IV-curve versus thickness map (f,~l). The black vertical lines at the $\Gamma$-point in the ARRES maps correspond with
the black vertical lines in the IV-curve map including left and right edge. All images have the same color intensity.
    }
\label{fig:ARRESthickness}
\end{figure*}

To fingerprint the difference between conducting and insulating samples at the growth temperature, we use
angle-resolved reflected electron spectroscopy (ARRES)~\cite{jobst15} as shown in Fig.~\ref{fig:ARREScompare}b-f.
ARRES maps of sample~\SampleCond{}, \SampleIns{} and \SampleSrO{} are shown in Fig.~\ref{fig:ARREScompare}b, d and e
respectively. These maps were measured directly after growth, at the growth temperature. The differences between the
conducting and the non-conducting samples are large. The conducting sample~\SampleCond{} (Fig.~\ref{fig:ARREScompare}b)
shows a band (minimum in intensity) around 14~eV at the $\Gamma$-point and a V-shaped band at the top of the figure
above 20~eV, while the insulating samples~\SampleIns{} and \SampleSrO{} (Fig.~\ref{fig:ARREScompare}d,e) show a maximum
(\ie{} a band gap) between 14 and 22~eV around the $\Gamma$-point.

In order to see whether this correlation is general, we measured two samples grown in other systems in ways which are
known from literature to produce conducting and non-conducting samples. Sample~\SampleTwente{} was grown in a
conventional PLD system with the possibility to grow under higher oxygen pressures which is known to result in
conducting samples. Sample~\SampleSputtered{} was grown by on-axis sputter deposition, known to result in insulating
samples~\cite{dildar13}. ARRES maps are shown in figure~\ref{fig:ARREScompare}c and f for \SampleTwente{} and
\SampleSputtered{} respectively. Their (non-)conductance is confirmed by electrical measurements
(Fig.~\ref{fig:ARREScompare}a). During the ARRES measurements, both samples were kept at a high temperature in an
oxygen pressure of $5\times 10^{-5}$~mbar to remove any contaminants and prevent the surface from charging. Exact
growth and measurement conditions can be found in the appendix. Comparing~\SampleCond{} and~\SampleTwente{} we conclude
the ARRES maps are similar and not sensitive to ex-situ transfer. The insulating samples~\SampleIns{}, \SampleSrO{}
and~\SampleSputtered{} in the bottom row of Fig.~\ref{fig:ARREScompare} are also similar, which leads us to conclude
that the differences are intrinsic.

Next we consider the change of the reflectivity with thickness. Fig.~\ref{fig:ARRESthickness} shows ARRES maps at the
growth temperature for every second unit cell grown. Conducting sample~\SampleCond{} is shown at the top
(Fig.~\ref{fig:ARRESthickness}a-e) and insulating sample~\SampleIns{} at the bottom (Fig.~\ref{fig:ARRESthickness}g-k).
Both samples start with a \TiO{}-terminated \STO{} surface (a,~g), showing the same map only slightly different in
brightness. The maps show a strong change as soon as two unit cells of \LAO{} are grown (b,~h). However, the maps of
the conducting sample~\SampleCond{} (top,~b) and insulating sample~\SampleIns{} (bottom,~h) still show many
similarities. This changes at four unit cells of \LAO{}. While for the conducting sample~\SampleCond{}
(Fig.~\ref{fig:ARRESthickness}c) the band around $\Gamma$ at 14~eV becomes a little bit more pronounced, the
non-conducting sample~\SampleIns{} (Fig.~\ref{fig:ARRESthickness}i) strongly changes and develops a pronounced band gap
around the $\Gamma$-point for energies between 14 and 22~eV, observed as a high-intensity area. Adding more \LAO{} up
to 6 (Fig.~\ref{fig:ARRESthickness}d,~j) and 8 (Fig.~\ref{fig:ARRESthickness}e,~k) unit cells only leads to little
changes, both for the conducting and the non-conducting samples.

\begin{figure*}[t]
\includegraphics[width=\textwidth]{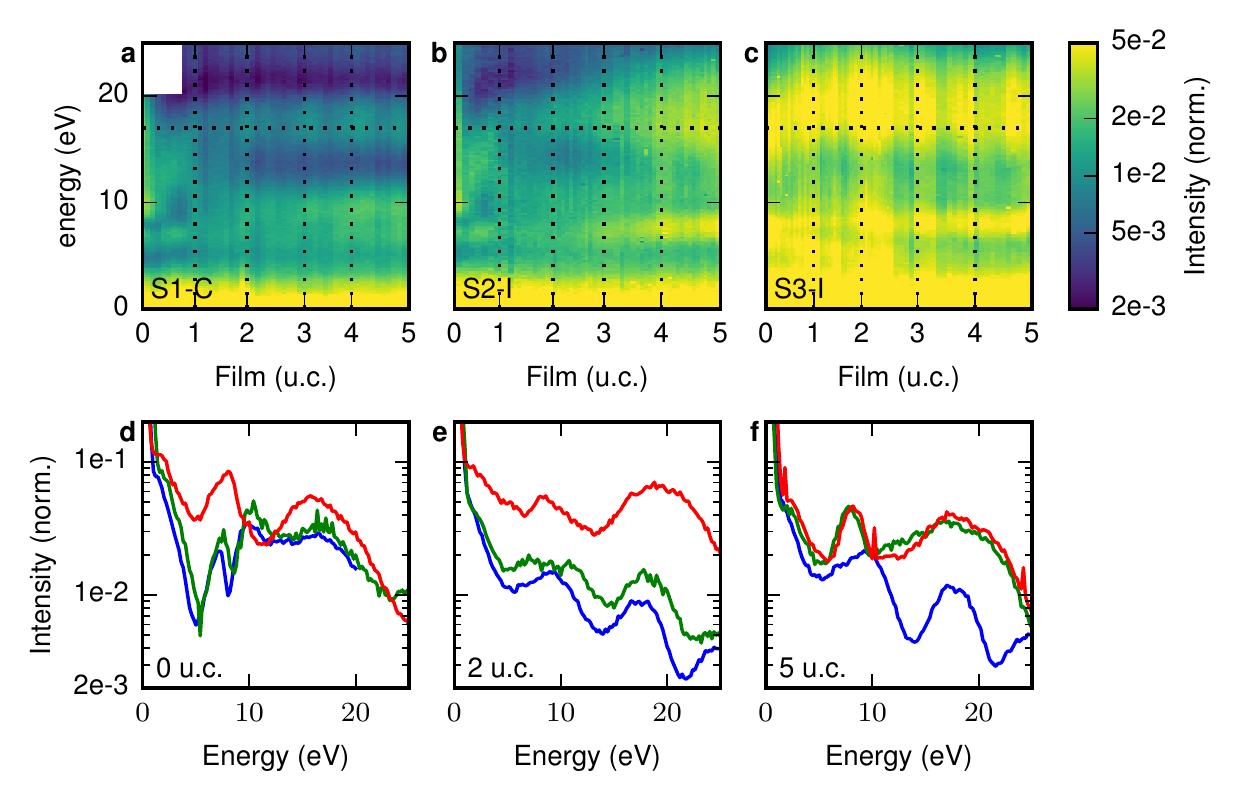}
\caption{ IV-curve versus thickness maps for sample~\SampleCond{} (a), sample~\SampleIns{} (b) and sample~\SampleSrO{}
(c). d, e, f) IV-curves after deposition of 0, 2 and 5 unit cells of \LAO{}, respectively, for sample~\SampleCond{}
(blue), sample~\SampleIns{} (green) and sample~\SampleSrO{} (red). Horizontal dotted lines in a,b and c indicate the
energy where Fig.~\ref{fig:FWHM} was measured. IV-curves are obtained by the integrated intensity of the specular
diffraction spot, filtering out any influence of the surface roughness. } \label{fig:IVthickness}
\end{figure*}

To probe the changes during growth in more detail we focus on the electron reflectivity at the $\Gamma$-point
($k_\parallel = 0$). This is nothing else than a LEEM (or LEED) IV-curve, which is the intensity variation of a
diffracted beam, in this case the specular beam, as function of electron energy. Such curves are indicated with a
vertical black line in the ARRES maps in Fig.~\ref{fig:ARRESthickness}. These curves were taken during growth at
regular intervals of 8 to 10 times per unit-cell. Results are shown in Fig.~\ref{fig:ARRESthickness}f and l
(sample~\SampleCond{} top (f) and sample~\SampleIns{} bottom (l)). They show the gradual change from the \STO{}
fingerprint to the final IV-curve of the \STO{}/\LAO{} heterostructure. The five solid black vertical lines between 0
and 8 u.c. correspond to the lines at the $\Gamma$-point in the five ARRES maps on the left side of
Fig.~\ref{fig:ARRESthickness}. The IV-curve map Fig.~\ref{fig:ARRESthickness}f shows that the band at 14~eV in
sample~\SampleCond{} appears just after two~unit cells have been grown. The band around 21~eV has already appeared at
this thickness. The non-conducting sample~\SampleIns{} (Fig.~\ref{fig:ARRESthickness}l) shows both bands around two
unit cells, but they vanish between three and four unit cells when the band gap appears between 14 and 22~eV. The band
gap at 8~eV also clearly appears at this thickness.

Still for samples \SampleCond{} and \SampleIns{} a zoomed-in part of the IV-curve maps for 0-5 unit cells is shown  in
figure~\ref{fig:IVthickness}a,b together with an IV-curve map of sample~\SampleSrO{} with \LAO{} on \SrO{}-terminated
\STO{} (Fig.~\ref{fig:IVthickness}c), the substrate prepared in a different PLD system. For comparison, the IV-curves
after deposition of 0, 2 and 5 unit cells of \LAO{} are plotted in figure~\ref{fig:IVthickness}d,e,f. Here the
IV-curves from sample~\SampleCond{} are plotted in blue, sample~\SampleIns{} in green and sample~\SampleSrO{} in red.
These plots clearly show two distinct IV-curves at 0 u.c. and two distinct IV-curves after deposition of 5~u.c.\@ of
\LAO{}. The starting IV-curves at 0 u.c. correspond with the \TiO{}- (blue, green) and \SrO{}-terminated (red) \STO{}
while in the IV-curves after deposition we distinguish the conducting (blue) and non-conducting (green, red) samples.
The evolution of the IV-curves during growth is different for the two insulating samples. This is very clear around 2
u.c. where sample~\SampleIns{} (green) is still close to sample~\SampleCond{} (blue) and not to sample~\SampleSrO{}
(red), which is already close to the insulating final IV-fingerprint found on the non-conducting samples. As a matter
of fact, the IV-curves for \SampleSrO{} hardly change during growth on the SrO-terminated surface.

With these results, we can return to Figure~\ref{fig:FWHM}, where for sample~\SampleCond{} the intensity strongly
decreased at the start and continued to oscillate around a low value; for sample~\SampleIns{} the intensity decreased
at the start, but recovered between 3 and 5 unit cells; and for sample~\SampleSrO{} the intensity oscillated around the
starting value, and did not decrease at all. The energy of 17~eV where the data of Fig.~\ref{fig:FWHM} was taken is
indicated with a horizontal dotted line in the IV-curve maps, Fig.~\ref{fig:IVthickness}a-c. Note that in
Fig.~\ref{fig:FWHM} the maximum of the specular diffraction spot is plotted, which is sensitive to spot broadening due
to surface roughening. This results in growth oscillations superimposed on the electron reflectivity signal. On the
other hand, for Fig.~\ref{fig:IVthickness} the intensity of the total specular spot is integrated, resulting in  an
intensity independent of spot shape (\ie{} surface roughness) and only depending on the electron reflectivity.
Combining Fig.~\ref{fig:FWHM} and~\ref{fig:IVthickness} we can now conclude that the increasing background signal
between 3 and 4 unit cells in Fig.~\ref{fig:FWHM}c is caused by the appearance of the band gap (enhanced surface
reflectivity) shown in Fig.~\ref{fig:IVthickness}b.

\begin{figure}[t]
\includegraphics[width=9cm]{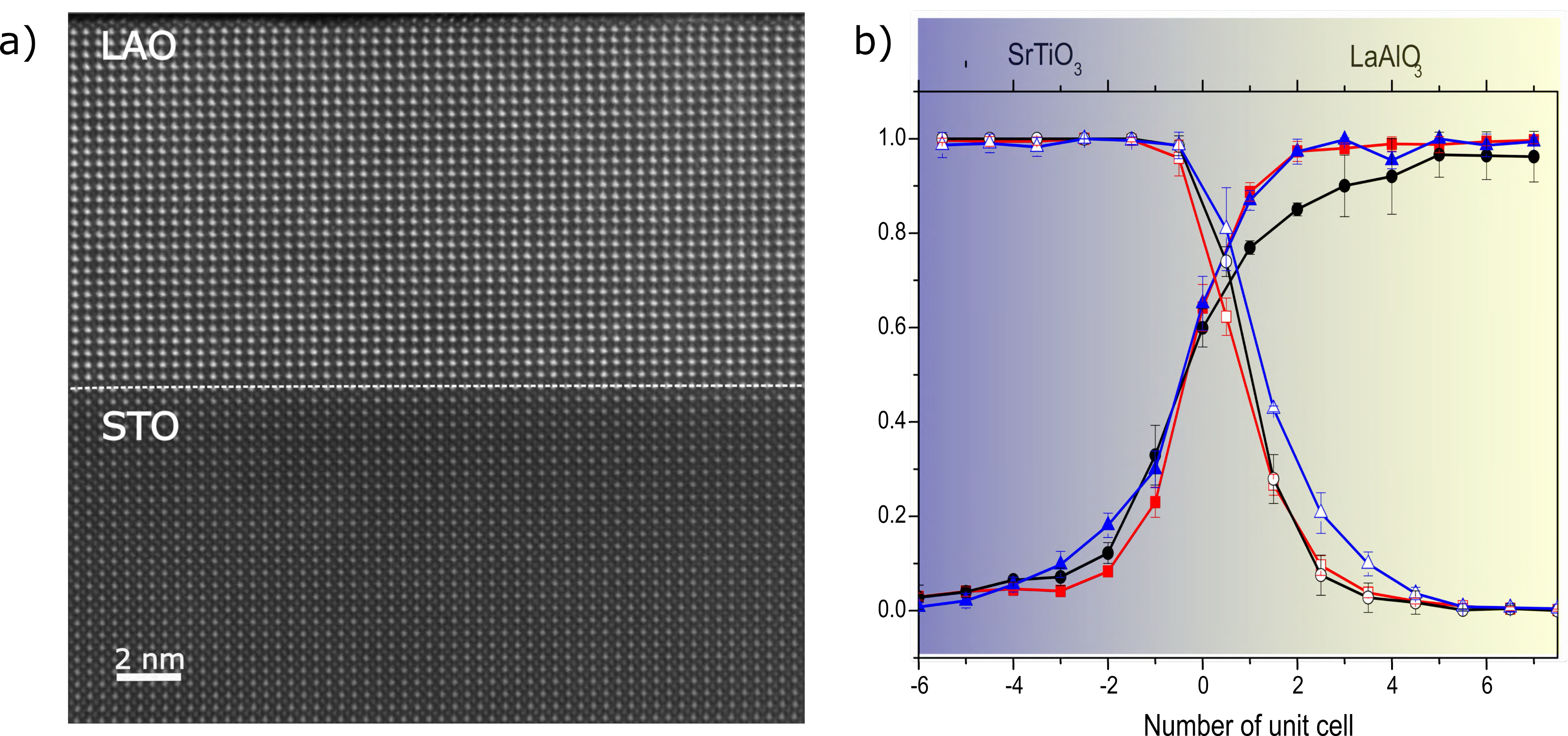}
\caption{
a) STEM-HAADF image in the [100] orientation of sample S6-I grown under the same conditions as S2-I but with 20 unit cells of \LAO{}.
Images for samples S7-C and S8-I show the same epitaxial quality. b) Normalized La- (filled symbols) and Ti-(open
symbols) occupancies for S6-I (blue triangles); S7-C (red squares) and S8-I (black circles) obtained from
EELS measurements.
}
%
\label{fig:STEM}
\end{figure}

One question with respect to the out-of-focus grown sample is whether the epitaxy is impaired by the ill-defined
fluence. For that we performed scanning transmission electron microscopy (STEM) experiments with high-angle annular
dark-field imaging (HAADF).
%
%
Samples S6-I, S7-C and S8-I were prepared in the same conditions as samples S2-I, S4-C and S3-I. For the sake of
avoiding any surface influence on the compositional analysis, thick films of 20UC were grown. Figure~\ref{fig:STEM}a
presents a typical STEM-HAADF of the \LAO{} film on the \STO{} substrate. Besides a slight misorientation of the
non-conducting films with respect to the substrate, good quality epitaxial growth was observed for all samples.
Figure~\ref{fig:STEM}b shows the La- and Ti-occupancies normalized to the total A- and B-site occupancy for samples
S6-I (defocused), S7-C and S8-I (SrO-terminated) in blue triangles, red squares and black circles, respectively. The
A-site is represented by filled symbols, the B-site by empty symbols. A similar extent of Ti-diffusion into the \LAO{}
(4-5 unit cells) can be observed for all samples, including the out-of-focus sample~S6-I. As expected, the
concentration of Ti in those first unit cells of the LAO film is higher, reflecting the Al-deficiency of the growing
film. On the other hand, The La/Sr intermixing is similar for samples S6-I and S7-C but Sr diffuses much further for
sample S8-I (SrO-terminated), leading to a relatively lower La content. We can therefore conclude that the A-site
interdiffusion is controlled by the substrate termination and not influenced by the out-of-focus condition.


\section{Discussion}
As mentioned in the introduction, the model of electronic reconstruction of a basically perfect interface is not enough
to explain the occurrence of conductance. Questions then exist about the relative importance of the role of
intermixing, oxygen vacancies, strain gradients with their ensuing buckling of the oxygen octahedra at the interface,
or the stoichiometry of the \LAO{} layer. The discussion on the La/Al-stoichiometry has started relatively recently. It
has been found that the \LAO{} film has to be Al-rich for conductance to appear~\cite{warusa13,brecken13a}, and also
that the \LAO{} stoichiometry is strongly dependent on the PLD parameters~\cite{brecken12}. We will now argue that our
electron reflectivity experiments precisely address the issues of stoichiometry and defects, which are crucial for the
occurrence of interface conductivity. Our observations are that (i) the difference between C- and I-samples is already
apparent during growth and at the growth temperature and (ii) the differences between C- and I-samples are significant
on the scale of eV’s. The conclusion we draw from this is that the (electronic) structure of the \LAO{} surface layer,
which is what our experiment is most sensitive to, is different for C-samples and for I-samples. The sensitivity of the
electron reflectivity to the surface layer can be demonstrated from the strong change in IV-curve seen in
Fig.~\ref{fig:IVthickness}d between \TiO{}-terminated and \SrO{}-terminated \STO{}. We note that the sensitivity
depends on the penetration depth, which is energy dependent. Unfortunately, calculations of the electron reflectivity
or the empty band structure of different possible surfaces do not yet exist in the measured energy range. We can
however sketch a scenario which can be considered for such calculations.

We start with noting that the \LAO{} grown on \TiO{}-terminated \STO{} should be \AlO{}-terminated, while the \LAO{}
grown on \SrO{}-terminated \STO{} should be \LaO{}-terminated. We surmise that this difference in termination causes
the strong difference between the conducting sample~\SampleCond{} and the non-conducting sample~\SampleSrO{}.
%
%
Consider now sample \SampleIns{}, which shows an IV-curve comparable to sample~\SampleCond{} (\AlO{}-terminated) for 2
unit cells (Fig.~\ref{fig:IVthickness}e) but changes to the signature of sample~\SampleSrO{} (LaO-terminated) for 5
unit cells (Fig.~\ref{fig:IVthickness}f). In contrast to the other samples, sample~\SampleIns{} was grown with an
out-of-focus laser. From literature we know that changing the PLD parameters, in particular the fluence, changes the
stoichiometry of the grown film. Furthermore, we know that Al-rich \LAO{} results in a conducting interface and La-rich
\LAO{} in an insulating interface. From this we infer that sample~\SampleIns{}, grown with an out-of-focus laser, is
La-rich. In growing \SampleIns{}, growth on \TiO{}-terminated \STO{} first results in a \AlO{}-termination, as seen
after growth of 2 unit cells. Growing further, the La-excess slowly builds up, changing the surface to \LaO{}-rich. We
note that the Ti-intermixing into the \LAO{} found for sample~\SampleIns{} could compensate the Al-deficits in the
first unit cells, suppressing the effects of the La-excess in those cells. Here we should remark that DFT calculations
in Ref.~\cite{weiland15} showed that surfaces are not \AlO{}- or \LaO{}-terminated, but rather that \surfAl{} and
\surfLa{} are the stable surface terminations. This implies that the \AlO{} surfaces mentioned above are actually
\surfAl{} and the \LaO{} surface are \surfLa{}, which does not conflict with our results. On the contrary, the fact
that less La is required for the \surfLa{} and more Al for the \surfAl{} surface could stimulate the transition from a
\surfAl{} to a \surfLa{} surface for our La-rich sample~\SampleIns{}. All in all, we argue that the strong change in
electron reflectivity, which is correlated to the unoccupied band structure, depends on the surface termination. From
the importance of the surface for the interface conductivity as described in literature~\cite{xie10,pentcheva12} and
our findings we deduce that the excess La in the surface layer could be an essential ingredient in suppressing the
electron transfer to the interface. More research has to be done to investigate the exact mechanism. Finally, we note
that our La-rich and Al-rich surface signatures do not correspond with the IV-curves measured on bulk mixed ordered
terminated \LAO{} as reported before~\cite{torren15}. This can however be explained by the surface reconstructions
found on the bulk \LAO{} and the difference between bulk and strained thin films.

\section{Summary}
We have shown results of electron reflectivity experiments (ARRES) on conducting and insulating
\LAO{}/\STO{}-heterostructures during growth, at the growth temperature with sub-unit cell precision. We find distinct
signatures for the conducting and non-conducting samples independent of their growth conditions. In other words, the
electron reflectivity (ARRES) can predict {\it during growth} whether a sample will show conductivity.

We find that the two families of reflectivity curves (maps) can be assigned to the surface termination being either
\AlO{} or \LaO{}-rich. For samples with Al-rich \LAO{} the surface termination is directly coupled to the termination
of the \STO{}. A \SrO{}-termination results in a \LaO{}-rich surface, while a \TiO{}-termination results in an
\AlO{}-rich surface. For the growth of La-rich \LAO{}, which we believe we achieve by out-of-focus laser growth, we
find the surface termination slowly changes from \AlO{}-rich to \LaO{}-rich during growth. From the importance of the
surface for the interface conductivity as described in literature\cite{xie10,pentcheva12}, we infer that it could be
this change in surface termination that is essential in suppressing the interface conductivity for the La-rich growth.

\section{Acknowledgements}
We want to acknowledge Ruud Tromp, Daniel Geelen, Johannes Jobst, Regina Dittmann, Gert Jan Koster, Guus Rijnders and
Jo Verbeek for discussions and advice and Ruud van Egmond and Marcel Hesselberth for technical assistance. This work
was supported by the Netherlands Organization for Scientific Research (NWO) by means of an ”NWO Groot” grant and by the
Leiden-Delft Consortium NanoFront. The work is part of the research programmes NWOnano and DESCO, which are financed by
NWO. N.G. acknowledges funding through the GOA project “Solarpaint” of the University of Antwerp and from the FWO
project G.0044.13N (Charge ordering). The microscope used in this work was partly funded by the Hercules Fund from the
Flemish Government. We would also like to acknowledge networking support by the COST Action MP 1308 (COST TO-BE).

\section{Appendix}

Five samples have been grown for LEEM analysis. The growth parameters of these films for PLD (S1-4) and sputtering (S5)
are shown in table~\ref{table:growth} together with the temperature where the ARRES maps are measured.

\begin{table}[h]
\begin{tabular}{llllll}
Sample Nr. 			& Fluence            & Growth    & Termination & Pressure        & ARRES\\
		 			& J/cm$^2$           & \dgC{}         &             & mbar          & \dgC{} \\
\SampleARRESCond{}	& 2					 & 780            & \TiO{}      & $5\times 10^{-5}$    & 795 \\
\SampleTwente{}		& 1					 & 720            & \TiO{}      & 1$\times 10^{-4}$    & 630 \\
\SampleOutOfFocus{} & defocus            & 770            & \TiO{}      & $5\times 10^{-5}$    & 770 \\
\SampleSrO{}		& 2                  & 700            & \SrO{}      & $5\times 10^{-5}$    & 600 \\
\SampleSputtered{}  & n.a.               & 830            & \TiO{}      & 3$\times 10^{0}$     & 560 \\
\end{tabular}
\caption{ PLD and sputter growth conditions for samples analyzed in LEEM as well as the temperature where the ARRES
maps were taken. } \label{table:growth}
\end{table}

\section*{References}
%

\end{document}